\documentclass[sigconf, nonacm, pdfa]{acmart}

\newcommand\vldbdoi{10.14778/3750601.3750701}
\newcommand\vldbpages{5493 - 5498}
\newcommand\vldbvolume{18}
\newcommand\vldbissue{12}
\newcommand\vldbyear{2025}
\newcommand\vldbauthors{\authors}
\newcommand\vldbtitle{\shorttitle} 
\newcommand\vldbavailabilityurl{https://ml-assets-management.github.io/}
\newcommand\vldbpagestyle{empty}

\usepackage{graphicx}
\usepackage{balance} 
\usepackage{amsmath,amsfonts}
\usepackage{textcomp}
\usepackage{latexsym}
\usepackage{color}
\usepackage{epsfig}
\usepackage{xspace}
\usepackage{pbox}
\usepackage{caption}
\usepackage{epsfig}
\usepackage{url}
\usepackage{mathrsfs}
\usepackage{booktabs}
\usepackage{adjustbox}
\usepackage{array}
\usepackage{multirow}
\usepackage{empheq}
\usepackage{tcolorbox}
\usepackage{stfloats}
\usepackage{enumitem}
\usepackage[colorinlistoftodos]{todonotes}
\usepackage{algorithm}
\usepackage{algpseudocode}
\usepackage{listings}
\usepackage{fbox}

\makeatletter
\DeclareRobustCommand*\cal{\@fontswitch\relax\mathcal}
\makeatother

\newcommand{\eat}[1]{}

\newcommand{\bi}{\begin{itemize}}
\newcommand{\ei}{\end{itemize}}
        {\end{itemize}\vspace{-1.5ex}}

\newcommand{\be}{\begin{enumerate}}
\newcommand{\ee}{\end{enumerate}}
\newcommand{\beqn}{\begin{eqnarray*}}
\newcommand{\eeqn}{\end{eqnarray*}}

\newcommand{\stitle}[1]{\vspace{.5ex}\noindent{\bf #1}}
\newcommand{\eetitle}[1]{\vspace{0.8ex}\noindent{\underline{\em #1}}}

\newcommand{\eg}{\emph{e.g.,}\xspace}

\newcommand{\kw}[1]{{\ensuremath{\mathsf{#1}}}\xspace}
\newcommand{\eop}{\hspace*{\fill}\mbox{$\Box$}}     
\newcounter{example}

\newcommand{\nthesection}{\arabic{section}}

\newcounter{prop}
\renewcommand{\theprop}{\arabic{theorem}}
\newcounter{lemma}

\newcounter{cor}
\renewcommand{\thecor}{\arabic{theorem}}
\newcounter{definition}[section]

\newcounter{alg}[section]
\renewcommand{\thealg}{\nthesection.\arabic{alg}}

\newcounter{arule}
\renewcommand{\thearule}{\arabic{arule}}

\newcommand{\ml}{\kw{ML}}
\newcommand{\ma}{\kw{ML} assets}
\newcommand{\mam}{\kw{ML}-asset management}
\newcommand{\mac}{\kw{ML}-asset curation}
\newcommand{\mad}{\kw{ML}-asset discovery}
\newcommand{\mau}{\kw{ML}-asset utilization}

\begin{document}
\title{\ml-Asset Management: Curation, Discovery, and Utilization}

\author{Mengying Wang}
\affiliation{%
  \institution{Case Western Reserve University}
  \city{Cleveland}
  \state{OH}
  \country{US}
}
\email{mxw767@case.edu}

\author{Moming Duan}
\affiliation{%
  \institution{National University of Singapore}
  \country{Singapore}
}
\email{moming@nus.edu.sg}

\author{Yicong Huang}
\affiliation{%
  \institution{University of California, Irvine}
  \city{Irvine}
  \state{CA}
  \country{US}
}
\email{yicongh1@ics.uci.edu}

\author{Chen Li}
\affiliation{%
  \institution{University of California, Irvine}
  \city{Irvine}
  \state{CA}
  \country{US}
}
\email{chenli@ics.uci.edu}

\author{Bingsheng He}
\affiliation{%
  \institution{National University of Singapore}
  \country{Singapore}
}
\email{hebs@comp.nus.edu.sg}

\author{Yinghui Wu}
\affiliation{%
  \institution{Case Western Reserve University}
  \city{Cleveland}
  \state{OH}
  \country{US}
}
\email{yxw1650@case.edu}

\begin{abstract}
Machine learning (\ml) assets, such as models, datasets, and metadata—are central to modern \ml workflows. 
Despite their explosive growth in practice, these assets are often underutilized due to fragmented documentation, siloed storage, inconsistent licensing, and lack of unified discovery mechanisms, making \mam\ an urgent challenge.
This tutorial offers a comprehensive overview of \mam\ activities across its lifecycle, including curation, discovery, and utilization. 
We provide a categorization of \ma, 
and major management issues, survey state-of-the-art techniques, and identify emerging opportunities at each stage.
We further highlight system-level challenges related to scalability, lineage, and unified indexing.
Through live demonstrations of systems,  
this tutorial equips both researchers and practitioners with actionable insights and practical tools for advancing \mam\ in real-world and domain-specific settings.
\end{abstract}

\maketitle

\pagestyle{\vldbpagestyle}
\begingroup\small\noindent\raggedright\textbf{PVLDB Reference Format:}\\
\vldbauthors. \vldbtitle. PVLDB, \vldbvolume(\vldbissue): \vldbpages, \vldbyear.\\
\href{https://doi.org/\vldbdoi}{doi:\vldbdoi}
\endgroup
\begingroup
\renewcommand\thefootnote{}\footnote{\noindent
This work is licensed under the Creative Commons BY-NC-ND 4.0 International License. Visit \url{https://creativecommons.org/licenses/by-nc-nd/4.0/} to view a copy of this license. For any use beyond those covered by this license, obtain permission by emailing \href{mailto:info@vldb.org}{info@vldb.org}. Copyright is held by the owner/author(s). Publication rights licensed to the VLDB Endowment. \\
\raggedright Proceedings of the VLDB Endowment, Vol. \vldbvolume, No. \vldbissue\ %
ISSN 2150-8097. \\
\href{https://doi.org/\vldbdoi}{doi:\vldbdoi} \\
}\addtocounter{footnote}{-1}\endgroup

\ifdefempty{\vldbavailabilityurl}{}{
\vspace{.3cm}
\begingroup\small\noindent\raggedright\textbf{PVLDB Artifact Availability:}\\
The source code, data, and/or other artifacts have been made available at \url{\vldbavailabilityurl}.
\endgroup
}

\section{Introduction}
\label{sec:intro}

\begin{figure}[tb!]
\centerline{\includegraphics[width = 0.46\textwidth]{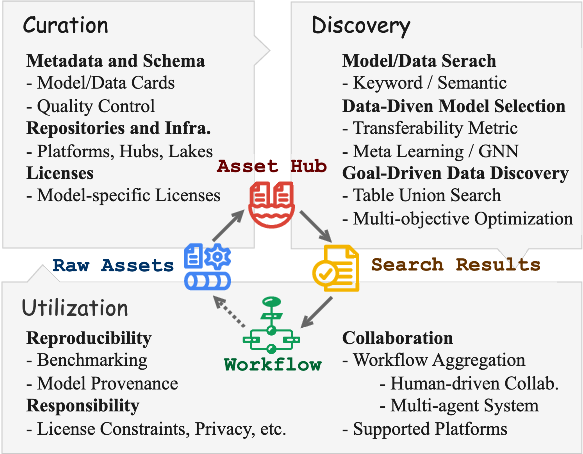}}
    \centering
    \caption{ML-Asset Management Lifecycle Overview}
    \label{fig:outline}
\end{figure}

The realm of Machine Learning (\ml) and Artificial Intelligence (AI) witnesses the creation of large amounts of \ml models and relevant data resources.
For example, data platforms such as HuggingFace~\cite{huggingface} host over $1.5$ million models, with $100,000$ new models added each month, occupying over $17$~PB of storage~\cite{horwitz2025charting}.
These are valuable {\em \ma}. 
Generally speaking, \ma\ are high-value artifacts that may contribute to ML-driven data analysis workflows.
Such \ma\ include, but not limited to:
\begin{itemize}[nosep, leftmargin=*, align=left]
    \item \textit{Datasets}: raw datasets, standardized documentation, annotated (training) data, validation data, test data, generated (benchmark) datasets, open data samples, feature vectors, etc; 
\item \textit{Models}: pre-trained, fine-tuned, or foundation ML models; and relevant supporting code resources such as training pipelines,  software libraries (e.g., AutoML components, LLM agents, statistical or physical models); 
\item \textit{Metadata}: ontologies or data constraints/rules; (open source or proprietary) licenses, scripts and prompts (\eg for LLMs), provenance data, data sources (\eg contributors),  hardware metadata, domain-specific in-lab/instrumental/experimental data, etc. 
\end{itemize}


The lack of management of rich sets of \ma\ leads to high maintenance costs, underutilized datasets and models, inefficiency in workflow development, and security and trustworthiness concerns. 
For example, over half of the models hosted in HuggingFace have no accompanying model card (documentation), and less than $8\%$ are properly licensed~\cite{horwitz2025charting}. 
Having this said, there still lack a standard characterization and through investigation of \mam\ issues. 
A cornerstone step is to establish a systematic characterization of \mam\ tasks and critical issues, and to provide a structured management infrastructure for modern \mam. 
Data management community plays an essential role in contributing fundamental and advanced data management techniques to support such needs -- by treating \ma\ with the same rigor as data objects, we can enable storage and modeling, queries and indexing, version control, lineage tracking, and provenance, among other critical capabilities.
\textbf{This tutorial} aims to provide an overview of major \mam\ tasks, synergetic research efforts that provide enabling techniques for effective \mam\ tasks, and a vision for future opportunities.

\stitle{Target Audience. } Our 
targeted audience comprises academic researchers, 
industry practitioners 
and stakeholders 
in both (1) data management, data science, 
machine learning, AI, and (2) multidisciplinary 
areas where ML-driven analysis plays critical roles.   

\stitle{Difference with Existing Tutorials. }
We structure our tutorial following a 
``life-cycle'' analysis of \ma, 
from the modeling and curation, 
resource access and discovery, 
to their utilization in downstream tasks. 
To ensure effective learning experience, 
we will incorporate a series of 
live, interactive demo to reveal 
the asset management techniques and 
their connections. 

\stitle{Learning Outcomes. }
Participants will gain a practical and conceptual understanding of \mam\ across, including

\begin{enumerate}[nosep, leftmargin=*, align=left]
    \item How to characterize and curate \ma\ effectively;
    \item How to store, index, and retrieve \ma\ at scale;
    \item How to manage provenance, versioning, and licensing; and 
    \item How can \mam\ support reproducibility, responsible, and collaborative AI?
\end{enumerate}


\vspace{-1ex}
\section{Tutorial Outline}
\label{sec:outline}

We propose a 1.5-hour tutorial to review the existing progress, challenges, and opportunities of \mam\ (outlined in Fig. \ref{fig:outline}). We next describe the detailed agenda for each part.

\begin{small}
\begin{figure}[tb!]
\begin{tcolorbox}[
    width=\linewidth,
    colback=white,
    colframe=black,
    fonttitle=\small,
    colbacktitle=gray!30,
    coltitle=black,
    sharp corners,
    boxrule=0.5pt,
    boxsep=5pt,
    left=1.5mm,
    right=1.5mm,
    top=0.5mm,
    bottom=1mm
]
\textbf{Part 1: Motivation and Background (5 mins)}

\textbf{Part 2: \ml-Asset Curation (22 mins)}
\begin{itemize}
    \item \textbf{Metadata and Schema}: Support Structured Understanding.
    \item \textbf{Repositories and Infrastructure}: Backbone of Discovery.
    \item \textbf{Licenses}: Enable Responsible Reuse.
\end{itemize}
\textbf{Part 3: \ml-Asset Search and Discovery (23 mins)}
\begin{itemize}
    \item \textbf{Model and Dataset Search.}
    \item \textbf{Data-driven Model Selection.}
    \item \textbf{Model-driven Data Discovery.}
\end{itemize}
\textbf{Part 4: \ml-Asset Utilization (25 mins)}
\begin{itemize}
    \item \textbf{Collaboration}: Workflow Aggregation and Automation.
    \item \textbf{Reproducibility}:  Benchmarking and Model Provenance.
    \item \textbf{Responsibility}:  Licensing and Ethical Asset Governance.
\end{itemize}
\textbf{Part 5: System Challenges and Opportunities (15 mins)}
\begin{itemize}
    \item \textbf{Storage and Scalability. }
    \item \textbf{Versioning and Lineage. }
    \item \textbf{Indexing and Searching. }
\end{itemize}
\textbf{Part 6: Demonstration (Split and merged into Part 2-4)}
\end{tcolorbox}
\caption{Tutorial Outline (90 minutes)}
\label{fig:outline}
\end{figure}
\end{small}

\subsection{\ml-Asset Curation}
\label{sec:modeling}

\mac\ is essential to harnessing the growing abundance of datasets and models.
Effective curation ensures that assets are not merely stored but also well-described, reusable, and regulated.

\eetitle{\textbf{Metadata and Schema}: Support Structured Understanding. } 
Metadata describes the essential properties of datasets and models, including their origin, modality, training configuration, evaluation metrics, known limitations, etc~\cite{akhtar2024croissant, li2023metadata}. 
It also captures interactions between assets, such as model performance across different datasets~\cite{koomthanam2024common}.
High-quality metadata helps clarify where an asset works well and where it might not, thereby minimizing misuse.

Community efforts have introduced concepts like Data Cards~\cite{pushkarna2022data} and Model Cards~\cite{mitchell2019model} to add structures to meta-information of \ma. 
One step forward, recent work such as CRUX~\cite{wang2022crux} provide 
knowledge graphs of \ma\ by linking various types of ``Cards'' (structured 
data objects).
Efforts on linking \ma\ with data dependencies, license compatibility, and 
functional calls among \ma\ with graph models opens the door 
for formal modeling, schema, and normalization, declarative manipulation, provenance, compatibility checking, among other 
issues that are of both theoretical and practical interests. 

\eetitle{\textbf{Repositories and Infrastructure:} Backbone of Discovery. }
Open platforms such as Hugging Face~\cite{huggingface}, Kaggle~\cite{kaggle}, TensorFlow Hub~\cite{TensorFlowHub}, and OpenML~\cite{openml} have become community standards for hosting ML models and datasets. They offer structured metadata templates, tagging, version control, and integration with popular ML frameworks for easy access and reuse.
More recently, the emergence of Data Markets~\cite{fernandez2020data}  and Model Lakes~\cite{ pal2024model} reflects a shift toward infrastructure that supports scalable discovery and reuse of \ma. These repositories move beyond siloed storage, enabling asset-centric querying, composition, and integration at scale.

\eetitle{\textbf{Licenses}: Enable Responsible Reuse. }
A unique aspect of \ma\ is the emergence of model-specific licenses, which are designed to govern the use and distribution of model-related components such as weights, checkpoints, optimizer states and architectures through legal terms and agreements.
Model-specific licenses introduce three key differences compared to traditional software licenses~\cite{rosen2005open}: governance over remote access (e.g., Model-as-a-Service~\cite{la2024language}), restrictions on responsible AI use~\cite{contractor2022behavioral} and conditions pertaining to model distillation and generated content.
For example, Gemma License~\cite{gemma2024} includes web access in its definition of ``Distribution'' and states that the ``transfer of patterns of the weights'' constitutes ``Model Derivatives'' governed by this license. 
Additionally, Article 3.1 of Gemma License prohibits licensees from using the Gemma model or its ``Model Derivatives'' in violation of its Prohibited Use Policy.
These tailored terms in model-specific licenses significantly broaden the scope of governance and increase 
curated objects, thereby complicating legal compliance in \ml systems.

There are several \textit{challenges and open opportunities} in \mac.
First, \textbf{incomplete or low-quality metadata}, much of the published metadata is manually entered without validation or quality control~\cite{neumaier2016automated}, which harms metadata-based discovery. One promising approach is automating metadata generation~\cite{cardinaels2005automating, schelter2017automatically} and utilizing LLMs for enhancing semantics. 
Second, \textbf{schema inconsistency}: different platforms adopt different metadata schemas, making integration and federation difficult. A community-wide standard schema or ontology is necessary to facilitate widespread adoption~\cite{guha2016schema}.
Third, \textbf{license ambiguity and conflicts}, for example, license restrictions can propagate through model fine-tuning chains, and some may be mutually exclusive (e.g., GPL-3.0 and Llama3.1 Community License~\cite{FSF2025various}). 
As model dependencies grow deeper and involve more components, manual legal compliance analysis becomes increasingly impractical~\cite{duan2024modelgo, duan2024they}. A formal license curation framework that enables automated dependency reasoning and compliance checking across assets is an essential next step.

\subsection{\ml-Asset Search and Discovery}
\label{sec:discovery}

Effective search and discovery are critical for the reuse of \ma\ and the acceleration of workflow construction~\cite{pei2023data}. 

\eetitle{\textbf{Model and Dataset Search. }}
Beginning with \textbf{keyword and tag-based filtering}, which is widely used on platforms such as Hugging Face~\cite{huggingface} and Kaggle~\cite{kaggle}, these systems support faceted search over structured metadata, enabling users to refine results based on modality, task, or license through exact matches~\cite{chapman2020dataset}. 
Recent progress has brought about \textbf{semantic and vector-based retrieval}, embedding models or datasets within a unified space for similarity-based searches~\cite{klabunde2023similarity}, with vector databases used to index model and document embeddings for rapid similarity queries~\cite{wang2021milvus}.
These methods offer entry points for \mad, but they tackle various asset types independently and neglect valuable interactions.

\eetitle{\textbf{Data-driven Model Selection.}}
Given high-value datasets and tasks at hand, a crucial question is: \textit{Which model should we use?} Brute-force evaluation is often infeasible due to the scale of modern model hubs~\cite{sparks2015automating,kumar2016model}. 
To address this, several works propose \textbf{transferability metrics}, which rank pre-trained models by estimating the label evidence on a target dataset, based on features extracted from the models~\cite{nguyen2020leep, you2021logme}. 
Another direction leverages \textbf{meta-learning} for model recommendation, where a recommender is trained to predict model performance using the metadata of \ml-assets. This approach enables more precise and context-aware ranking~\cite{kotlar2021novel}.
\textbf{Graph learning-based} recommenders may further improve the quality of 
suggested models, by exploiting enriched metadata/features, 
better-informed suggestions and annotations, and more efficient cold-start 
strategies for new datasets~\cite{wang2023selecting, li2024model}.

\eetitle{\textbf{Model-driven Data Discovery.}} 
Recent research also advocates that
data discovery for \ml models could be ``model-driven'', 
with a goal to identify data over which 
a given \ml model has high expected performance and 
small training/testing overhead. 
This requires finer-grained data manipulation that may integrate  
 feature engineering and data integration~\cite{li2024amalur}. 
Methods utilizing \textbf{table union search} to enhance data completeness and schema compatibility may generate tables by semantically merging  
multiple contextualized columnar data sources~\cite{fan2023semantics}.
\textbf{Goal-oriented data discovery} tailors data selection to specific downstream tasks, guided by a target utility function~\cite{galhotra2023metam}. 
\textbf{Multi-objective data discovery} incorporates multiple user-defined model performance evaluation criteria, to generate datasets that may optimize 
model performance across various performances~\cite{wang2025generating}.
Recent research also develop model-aware data augmentation for LLM pretraining and fine-tuning~\cite{kangget, wang2024llm, xie2023data}.

Effective \mad\ benefits from guarantees on \textit{robust search}, \textit{high-quality metadata}, and \textit{context-awareness}. Yet, several challenges remain alongside opportunities. 
First, \textbf{cold-start issue}: current methods depend heavily on underlying metadata, which might be limiting for many tasks. Beyond enriching metadata, embedding techniques to derive standardized representations directly from raw assets content offers a promising solution.
Second, \textbf{discovery at scale}, searching through huge number of assets with
complex queries becomes computationally expensive. Scalable infrastructure (distributed retrieval,
caching of embeddings, vector databases, etc.) is needed.
Third, \textbf{semantic understanding}: interactive discovery requires systems to understand both sides (model and data) at a semantic level, making a unified representation space that encapsulates the characteristics of various asset types crucial.

\subsection{\ml-Asset Utilization}
\label{sec:opportunities}

The ultimate payoff lies in utilizing these well-organized \ma\ and supporting systems to improve applications and practices in \textit{reproducible}, \textit{ethical}, and \textit{collaborative} data science~\cite{arnold2019turing}.

\eetitle{\textbf{Collaboration}: Workflow Aggregation and Automation.}
Workflow aggregation involves creating modular ML workflows by selecting compatible assets from repositories. 
This modularity enables collaboration~\cite{liu2022demonstration}, allowing teams from different disciplines and backgrounds to collaboratively integrate and reuse components from various sources~\cite{derakhshan2020optimizing}.
Platforms such as Davos~\cite{shang2021davos} and Texera~\cite{wang2024texera} have emerged to support these efforts.
Modularized workflows can be modeled as directed acyclic graphs (DAGs), providing a structure foundation for aggregation approaches~\cite{sparks2017keystoneml}.
Beyond human-driven collaboration, there is a trend toward automating workflow construction using agents.
Unlike traditional AutoML, which typically requires extensive testing~\cite{nakandala2020cerebro}, multi-agent approaches frame workflow assembly as a goal-conditioned planning problem.
Leveraging language agents, 
such a framework may reason over asset metadata, infer task requirements, and iteratively assemble workflows~\cite{zhang2025aflow}.

\eetitle{\textbf{Reproducibility}: Benchmarking and Model Provenance.}
\mam\ streamlines benchmarking, allowing researchers to evaluate algorithms against standard datasets and baselines stored in asset repositories~\cite{longjohn2024benchmark}.
Employing established procedures by leveraging version-controlled and validated resources significantly improves the reproducibility of experiments~\cite{thiyagalingam2022scientific}.
A second critical aspect is tracking model provenance~\cite{namaki2020vamsa,psallidas2023demonstration}, which aims to track a model's lineage data (training data, preprocessing, hyperparameters, source code, evaluation metrics and results).
Model provenance allows others to replicate experiments, verify reported results, and gain insights in their reusability~\cite{rupprecht2020improving}. 
Data provenance such as ``Why-provenance'' can be adapted to generate post-hoc explanations for tracking ML model outputs, as observed in~\cite{chen2024view,qiu2024generating}.

\eetitle{\textbf{Responsibility}: Licensing and Ethical Asset Governance.}
Responsible \mau\ starts with clear licensing and agreements that govern how an asset may be used, adapted, or distributed. 
Licensing information often suffers from inconsistency during the reuse and republishing of licensed materials~\cite{wu2017analysis}.
As an open standard for \textit{AI Bills of Materials} (BOM), \textit{Software Package Data Exchange 3.0} (SPDX 3.0)~\cite{karen2024implementing} enables the structured recording of \ma\ and their associated licensing information throughout the development lifecycle, potentially supporting automated license-related analyses, such as detecting compatibility issues~\cite{kapitsaki2017automating}, inconsistencies~\cite{wu2015method, wu2017analysis}, and license proliferation~\cite{gomulkiewicz2009open}.
Existing license compliant analysis tools such as FOSSology~\cite{jaeger2017fossology}, Carneades~\cite{gordon2011analyzing}, ModelGo Analyzer~\cite{duan2024they} and Black Duck~\cite{blackduck2024} may extend to ML projects if AI BOM is avaliable.
Unfortunately, SPDX 3.0 is not yet integrated into mainstream ML tools, and model development disclosure remains unstandardized.
Moreover, commonly used model file formats (e.g., Safetensors, GGUF, and OpenVINO IR) do not embed license metadata, leading to inconsistency of licensing information. 
\mau\ remains subject to uncertain legal compliance risks.

Beyond licensing, responsible reuse also involves privacy and transparency issues which are expected to be captured through metadata documentation (e.g., model/data cards and provenance data).
However, this remains challenging due to the variation in AI-related regulations across jurisdictions. 
Meanwhile, the use policies of the model vendors (enforceable under contract law) must also be complied with. 
Material breaches of either applicable laws or licenses/agreements may result in legal consequences.



\subsection{System Challenges \& Opportunities}
\label{sec:challanges}

Treating \ma\ as first-class citizens indicates new types of \mam\ systems. 
As \ma\ grow rapidly in size, complexity, and volume, there is a need to revisit data management systems on 
how to best explore them in \mam. 

\eetitle{\textbf{Storage and Scalability}. }
\ma, particularly large models and datasets, pose substantial storage challenges due to their rapidly increasing size and complexity. 
Techniques have emerged as scalable solutions, such as \textbf{compressed binary formats} like Safetensors provide safe, zero-copy tensor storage, enabling faster and secure model loading during deployment~\cite{casey2025empirical}. Earlier systems like ModelDB adopted a lightweight design by \textbf{storing only essential metadata} while keeping large binaries in external object storage\cite{vartak2016modeldb}. Other efforts, such as Model Lake, leverage \textbf{distributed storage infrastructure}\cite{garouani2024model}, etc.
Ensuring efficiency, security, and consistency at scale remains an open research challenge. 

\eetitle{\textbf{Versioning and Lineage}. }
Versioning and lineage tracking are crucial for reproducibility and auditability in data science.
\textbf{Delta-based version control systems}, inspired by Git, allow efficient management of evolving datasets and models by capturing fine-grained changes along with detailed metadata~\cite{miao2017modelhub}. 
\textbf{Provenance systems} like ProvDB represent ML workflows as graphs, enabling rich queries over asset lineage and dependencies~\cite{miao2017provdb}. 
Nonetheless, scalability remains a considerable challenge, and the potential for lineage reuse is an area that warrants further investigation~\cite{phani2021lima}.

\eetitle{\textbf{Indexing and Searching}. }
To efficiently search through a large volume of assets, powerful indexing mechanisms are essential.
Supporting heterogeneous information, such as structured metadata, graph-based lineage, and semantic embeddings, poses challenges for hybrid query execution.
Early attempts often relied on interfaces that surfaced all data types separately (e.g., via tabbed views)~\cite{garouani2024model}, but hybrid indexes offer a more unified and performant solution~\cite{patel2024acorn}.
Additionally, it is important to maintain index freshness with minimal cost, while also considering privacy and security concerns. 
There are several potential directions: 
(1) Explore \textbf{vector database} systems~\cite{pan2024survey} 
and optimization techniques 
in vector processing for large-scale 
\ml-asset search. (2) 
Text-rich domain languages, datahubs, and application 
scenarios of \ma\ continue to enrich their metadata and features, hence in turn providing opportunities of recent Large Langage Models (\textbf{LLMs}) and Retrieval Augmented Generation (\textbf{RAG}) methods in \ml-asset recommendation.

\section{Demonstration}
\label{sec:demo}

We will walk through several \mam\ tools and showcase how they may benefit \mam\ tasks. 

\eetitle{\textbf{CRUX}: ML-Asset Curation and Discovery. }
CRUX~\cite{wang2022crux} is a 
crowdsourced platform for curating \ma\ analysis for materials data science. 
It captures rich metadata and model–dataset interactions using domain-specific ontologies co-designed with materials scientists.
CRUX supports
model recommendation and data discovery~\cite{wang2024modsnet}. We will demonstrate asset ingestion, metadata visualization, and search over asset knowledge graphs. 

\eetitle{\textbf{ModelGo}: \ml-Asset License Analyzer and  License set.} We demonstrate 
the following: ModelGo Analyzer~\cite{duan2024they, duanposition}, an ontology-based tool for automated license compliance analysis in \ml projects. It evaluates licensing-related issues such as rights granting, term conflicts, and incompatibility between licenses.
ModelGo Licenses~\cite{modelgoli2025}: a new Creative Commons-style model-specific license set designed for general model publication. It supports flexible licensing options to meet diverse model sharing needs. 

\eetitle{\textbf{Texera}: Collaborative Workflow Composition.}
Texera~\cite{wang2024texera, texera:website} is an open-source platform designed to support collaborative data science and AI/ML.
It offers a GUI-based workflow interface that enables analysts with diverse technical backgrounds to contribute effectively. 
Analysts can collaboratively edit workflows, interact with live executions~\cite{journals/pvldb/KumarWNL20}, and jointly debug a workflow execution~\cite{journals/pacmmod/HuangWL23} in real time.
Texera also supports reproducibility and determinstic replays~\cite{ni2024icedtea} by preserving execution configurations and histories.

\section{Biography}
\label{sec:bio}

\textbf{Mengying Wang} is a Ph.D.~candidate in Computer Science at Case Western Reserve University (CWRU), 
advised by Dr. Yinghui Wu. 
Her research interests include \mam\, agentic workflow, knowledge graph discovery, and graph RAG. 
\textbf{Moming Duan} is a Research Fellow at Institute of Data Science, National University of Singapore (NUS). His research interests include AI Governance and Licensing and Federated Learning.
\textbf{Yicong Huang} is a Ph.D.~candidate in the Department of Computer Science, University of California, Irvine (UCI), advised by Dr.~Chen Li. 
His research interests lie in data management and ML systems.
He is a main contributor of the Texera project.
\textbf{Chen Li} is a Professor in the Department of Computer Science at UCI.
His research interests are in data management, including data-intensive computing, databases, query processing, ML systems, and data science.
His current focus is building open-source systems for big data and AI/ML.
\textbf{Bingsheng He} is a Professor at School of Computing, NUS. His current research interests include database and machine learning systems, and high performance computing.
\textbf{Yinghui Wu} is an Associate Professor in the Department of Computer and Data Sciences, CWRU. His area is in data management, data science, and graph data analysis. 

\begin{acks}
Wang and Wu are supported by NSF under OAC-2104007. 
Huang and Li are supported by NSF under award III-2107150 and NIH under award 1U01AG076791-01.
Duan and He are supported by the National Research Foundation, Singapore and Infocomm Media Development Authority under its Trust Tech Funding Initiative.
\end{acks}


\bibliographystyle{ACM-Reference-Format}
\balance
\bibliography{paper}

\end{document}